\documentclass[aps, prb, reprint, showpacs, showkeys, onecolumn, 11pt, superscriptaddress, notitlepage]{revtex4-1}
\usepackage{graphicx} 
\usepackage{braket}
\usepackage{color}
\usepackage{times}
\usepackage{textcomp} 

\begin{document}

\title{Manipulation of the nuclear spin ensemble in a quantum dot with chirped magnetic resonance pulses}

\author{Mathieu Munsch}
\affiliation{Department of Physics, University of Basel, Klingelbergstrasse 82, CH-4056 Basel, Switzerland}

\author{Gunter W\"{u}st}
\affiliation{Department of Physics, University of Basel, Klingelbergstrasse 82, CH-4056 Basel, Switzerland}

\author{Andreas V. Kuhlmann}
\affiliation{Department of Physics, University of Basel, Klingelbergstrasse 82, CH-4056 Basel, Switzerland}

\author{Fei Xue}
\affiliation{Department of Physics, University of Basel, Klingelbergstrasse 82, CH-4056 Basel, Switzerland}

\author{Arne Ludwig}
\affiliation{Department of Physics, University of Basel, Klingelbergstrasse 82, CH-4056 Basel, Switzerland}
\affiliation{Lehrstuhl f\"{u}r Angewandte Festk\"{o}rperphysik, Ruhr-Universit\"{a}t Bochum, D-44780 Bochum, Germany}

\author{Dirk Reuter}
\affiliation{Lehrstuhl f\"{u}r Angewandte Festk\"{o}rperphysik, Ruhr-Universit\"{a}t Bochum, D-44780 Bochum, Germany}
\affiliation{Department Physik, Universit\"{a}t Paderborn, Warburger Strasse 100, D-33098 Paderborn, Germany}

\author{Andreas D. Wieck}
\affiliation{Lehrstuhl f\"{u}r Angewandte Festk\"{o}rperphysik, Ruhr-Universit\"{a}t Bochum, D-44780 Bochum, Germany}

\author{Martino Poggio}
\affiliation{Department of Physics, University of Basel, Klingelbergstrasse 82, CH-4056 Basel, Switzerland}

\author{Richard J. Warburton}
\affiliation{Department of Physics, University of Basel, Klingelbergstrasse 82, CH-4056 Basel, Switzerland}

\date{\today}

\begin{abstract}
The nuclear spins in nano-structured semiconductors play a central role in quantum applications \cite{Ribeiro2013,Warburton2013,Chekhovich2013,Greilich2007}. The nuclear spins represent a useful resource for generating local magnetic\cite{Urbaszek2013} fields but nuclear spin noise represents a major source of dephasing for spin qubits \cite{Warburton2013,Chekhovich2013}. Controlling the nuclear spins enhances the resource while suppressing the noise. Nuclear magnetic resonance (NMR) techniques are challenging: the group-III and group-V
isotopes have large spins with widely different gyromagnetic-ratios; in strained material there are large atom-dependent quadrupole-shifts \cite{Chekhovich2012};
nano-scale NMR is hard to detect \cite{Staudacher2013,Mamin2013}. We report NMR on $100,000$ nuclear spins of a quantum dot using chirped radio-frequency pulses. Following
polarization, we demonstrate a reversal of the nuclear spin. We can flip the nuclear spin back-and-forth a hundred times. We demonstrate that
chirped-NMR is a powerful way of determining the chemical composition, the initial nuclear spin temperatures and quadrupole frequency distributions
for all the main isotopes. The key observation is a plateau in the NMR signal as a function of sweep-rate: we achieve inversion at the first quantum transition for all isotopes simultaneously. These experiments represent a generic technique for manipulating nano-scale inhomogeneous nuclear spin ensembles and open the way to probe the coherence of such mesoscopic systems.
\end{abstract}

\maketitle

NMR signals can be boosted by polarizing the nuclei. This is particularly beneficial on the nano-scale where NMR signals are invariably small and hard to detect. The nuclear spins in a self-assembled quantum dot can be polarized optically by exploiting the hyperfine interaction with an electron spin \cite{Urbaszek2013,Chekhovich2013}. Extremely long-lived polarizations \cite{Greilich2007, Maletinsky2009, Latta2011b, suppl} up to about 50\% have been achieved. The nuclear spin polarization results in a shift of the optical resonance, the Overhauser shift, facilitating its sensitive detection \cite{Urbaszek2013}. These features have enabled the observation of isotope-selective NMR of the nuclear spins associated with strain-free GaAs quantum dots \cite{Gammon1997,Makhonin2011}. Self-assembled quantum dots, attractive for single photon generation and optically-controlled spin qubits \cite{Warburton2013}, have highly inhomogeneous nuclear spins \cite{Urbaszek2013,Flisinski2010,Cherbunin2011,Bulutay2012}. Additional side peaks appear in the NMR spectra, a consequence of a strain-dependent quadrupole interaction, along with a distribution of chemical shifts \cite{Chekhovich2012}. Manipulating the nuclear spin ensemble of a single quantum dot is challenging yet important: projection of the nuclear spins into a specific state boosts the single electron spin dephasing time \cite{Greilich2007}; developing techniques to probe nano-sized ensembles of highly inhomogeneous nuclear spins has impact also for semiconductor nanowires \cite{Peddibhotla2013} and nanocrystals. 

Here we use chirped NMR pulses. The main concept is that by sweeping over a large frequency range, the pulse addresses each nuclear spin at some point. For a spin-$\frac{1}{2}$ nucleus, a 2-level system, the Hamiltonian in the rotating frame is,
\begin{equation}
H=h \Delta \nu(t) I_{z} + \frac{1}{2} h \gamma B_{x} I_{x}
\end{equation}
where $h$ is the Planck constant, $I$ the nuclear spin, $\gamma$ the gyromagnetic ratio of the nuclear isotope (in frequency units) and $\Delta \nu(t)$ is the time-dependent detuning between the radio frequency (RF) excitation and the Larmor frequency $\nu_L = \gamma B_z$. The coupling between the RF magnetic field $B_{x}$ and the spin, the second term in the Hamiltonian, leads to an avoided crossing in the eigen-energies with splitting $h \nu_{RF}$, Fig.\ 1a , where $\nu_{RF}=\gamma B_{x}$. On traversing the avoided crossing from large and negative $\Delta \nu$ to large and positive $\Delta \nu$ with a single pulse ($N=1$) at sweep rate $\alpha$, the probability that the final state is $\ket{\uparrow}$ for initial state $\ket{\uparrow}$, is 
\begin{equation}
P_{LZ}=\exp(-\pi^2\nu_{RF}^{2}/\alpha),
\end{equation}
the Landau-Zener result \cite{Shevchenko2010}. In the sudden regime when $P_{LZ} \simeq 1$, the system ``tunnels" through the avoided crossing and $\ket{\uparrow} \rightarrow \ket{\uparrow}$, $\ket{\downarrow} \rightarrow \ket{\downarrow}$. Alternatively, in the limit when $P_{LZ} \ll 1$, the states are swapped $\ket{\uparrow} \rightarrow \ket{\downarrow}$, $\ket{\downarrow} \rightarrow \ket{\uparrow}$: this is adiabatic passage, Fig.\ 1a. 

We attempt to apply these concepts to a single nano-scale nuclear spin ensemble. The challenges are, first, each nuclear spin is more complex than a two-level system; and second, there is an inhomogeneous distribution of $10^{5}$ nuclear spins. Initialization and detection of the nuclear spin polarization of a single quantum dot is carried out optically with exquisite spectral resolution provided by resonant laser spectroscopy, representing a sensitivity to $\sim 1,000$ spins. The quantum dots for these experiments are gate-controlled In$_x$Ga$_{1-x}$As quantum dots\cite{suppl}, Fig.\ 1c. The bias voltage controls both the occupation of the quantum dot (here empty) and the exact optical transition frequency via the Stark effect. Key to reaching the adiabatic limit $P_{LZ}\ll 1$ is the generation of RF fields with high amplitude. We use an on-chip, low-impedance, high bandwidth microwire \cite{Poggio2007,suppl}, fabricated directly above the gate: large oscillating currents in the microwire generate oscillating magnetic fields ($B_{x} \simeq 5$ mT \cite{suppl}); the small impedance of the microwire enables fast pulsing. An aperture in the microwire allows optical access to the quantum dots directly underneath, Fig.\ 1b,d. The quantum dot optical resonance (X$^{0}$) is driven with a coherent laser with resonance fluorescence detection \cite{Kuhlmann2013a,Kuhlmann2013b}, the read-out after one RF pulse providing the initialization for the next, Figs.\ 1e and 2c. 

A resonance fluorescence spectrum of the quantum dot at zero applied magnetic field, $B_{z}=0$\,T, is shown in Fig.\ 2a: the two lines, split by the fine-structure, have linewidths of 1.2 \textmu eV, close to the transform limit of 0.9 \textmu eV  \cite{Kuhlmann2013c}. At $B_{z} \ge 0.5$ T, on sweeping through the optical resonance, the nuclear spins adjust their polarization to maintain an optical resonance of the quantum dot with the laser, the ``dragging" effect \cite{Latta2009,Hogele2012}: the Overhauser shift OHS equals the laser detuning $\delta_{L}$. Dragging represents a way of generating large bi-directional nuclear spin polarizations \cite{Latta2009}. An example is shown in Fig.\ 2b: starting with the nuclei in a depolarized state \cite{suppl}, the optical resonance is ``dragged" to $\delta_{L}=-41$ $\mu$eV. The nuclear spin polarization decays extremely slowly (timescale days for an empty quantum dot \cite{suppl, Greilich2007, Maletinsky2009, Latta2011b}), resulting in optical memory effects. A sequence of optical sweeps is shown in Fig.\ 2b: the rise point of each scan is related to the polarization set by the previous scan whereas the end of the plateau sets the new polarization state. For a given laser sweep direction, the change in width of the dragging ``plateau" following an NMR pulse is used to measure the change in the Overhauser field, $\Delta_{\rm OHS}$ in Fig.\ 2c. 

Manipulation of the nuclear spin ensemble is demonstrated in Fig.\ 2c. The nuclear spin polarization along $z$, $\left< I_{z} \right>$, is initialized with a sweep from positive to negative $\delta_{L}$. With the laser off, a chirped NMR pulse is applied, $\nu =\nu_{1} \rightarrow \nu_{2}$. The laser is then turned back on and the sweep from positive to negative $\delta_{L}$ repeated. The optical signal now appears not at negative $\delta_{L}$ but at positive $\delta_{L}$, unambiguous evidence that the RF pulse inverts the nuclear spin polarization. In this particular case, following optical polarization, $\left< I_{z} \right>/I_{z}^{\max} \simeq +32$\%, and after one NMR pulse,  $\left< I_{z} \right>/I_{z}^{\max} \simeq -13$\% \cite{suppl}. This interpretation is backed up by applying not one but a sequence of (phase-matched) chirped pulses, $\nu_{1} \rightarrow \nu_{2} \rightarrow \nu_{1} \rightarrow \nu_{2}$\ldots. As a function of pulse number $N$, $\left< I_{z} \right>$ oscillates from positive to negative, evidence of close-to-adiabatic manipulation of $\left< I_{z} \right>$. We can invert-restore the nuclear spin polarization $\sim 100$ times before the signal is lost, Fig.\ 2d.

We explore the dependence on sweep rate $\alpha$ on tuning from low $\nu_{1}$ to high $\nu_{2}$ such that all nuclear spins are addressed. The signal increases with decreasing sweep rate, Fig.\ 3. Significantly, there is an exponential increase followed by a plateau and then another exponential increase. The step-wise transition from the sudden to the adiabatic regime is a consequence of a hierarchy of avoided crossings in the energy level structure. It arises from a quadrupole interaction of the nuclear spin with a local electric field gradient resulting in an additional term in the Hamiltonian,
\begin{equation}
H_{Q}=\frac{1}{6} h \nu_{Q}\left[3 I_{z}^{2}-I(I+1) \right].
\end{equation}
where $h\nu_Q$ is the strength of the quadrupole field \cite{suppl}. Fig.\ 1a shows the eigen-energies for $I=\frac{3}{2}$, both for $\nu_{Q}=0$ and for $\nu_{Q} \gg \nu_{RF}$. When $\nu_{Q} \neq 0$, a hierarchy of avoided crossings appears, large for the first quantum transitions (bare states separated by $|\Delta m|=1$); intermediate at the second quantum transitions ($|\Delta m|=2$); and small at the third quantum transition ($|\Delta m|=3$). A similar but more complex hierarchy also arises in the In ($I=\frac{9}{2}$) eigen-energies. Given the exponential dependence of $P_{LZ}$ on the energy separation at the avoided crossing, this means that the different quantum transitions satisfy the adiabaticity condition at quite different sweep rates \cite{Vega1978, Haase1994, Veenendaal1998}. At the plateau in Fig.\ 3, the sweep is adiabatic for  the first quantum transitions ($P_{LZ} \ll 1$) whereas the others are still in the sudden regime ($P_{LZ} \simeq 1$). At first sight, it is surprising that the step signifying adiabaticity at the first quantum transitions survives the ensemble averaging. The explanation is to be found in the scaling of the energies at the avoided crossings, $h \nu_{\text{eff}}$. In the limit $\nu_{Q} \gg \nu_{RF}$, $\nu_{\text{eff}} \propto \nu_{RF} (\nu_{RF}/\nu_{Q})^{|\Delta m|-1}$ for all $I$ \cite{Veenendaal1998, Haase1994, Vega1978, suppl}. This means that for $|\Delta m|=1$, $\nu_{\text{eff}}$ does {\em not} depend on $\nu_{Q}$ (to first order), suppressing the sensitivity of the adiabaticity criterion to the quadrupole interaction. 

The plateau in the sweep rate dependence is the key observation that allows both the indium concentration $x$ and the initial nuclear spin temperature $T$ to be determined. The point is that the signal at the plateau, $\Delta_{\rm OHS}=$ 28.8 \textmu eV, and the initial Overhauser shift, ${\rm OHS}=$ 27.0 \textmu eV, are determined solely by $x$, $T$ and the known nuclear parameters (nuclear spins, hyperfine coupling constants and abundances of $^{75}$As, $^{115}$In, $^{69}$Ga and $^{71}$Ga) \cite{suppl}. We find $x=(20.2 \pm  5.7)$\% and $T=(8.2 \pm 0.8)$ mK. The composition $x$ represents the indium concentration over the extent of the electron wave function; the temperature, much lower than the bath temperature of 4.2 K, interprets the dynamic nuclear spin polarization as a laser cooling phenomenon. 

Spectroscopic identification of the isotopes is presented in Fig.\ 4 where the NMR pulse is chirped from a fixed $\nu_{1}$ to a variable $\nu_{2}$ using a slow and constant sweep rate. The NMR signal $\Delta_{\rm OHS}$ increases step-wise around 44 MHz. This arises when $\nu_{2}$ goes above the central NMR frequency of a particular isotope, in this case $^{75}$As. Another clear step arises at 79 MHz, the $^{71}$Ga resonance. Around the central transition, the single spin satellite steps \cite{suppl} are broadened through atom-dependent quadrupole couplings. This is particularly visible in the In contribution because of the large number of satellites. This curve enables us to determine the average quadrupole frequency $\left< \nu_Q \right>$ and an approximate distribution $p(\nu_{Q})$ for {\em all} the main isotopes, $^{75}$As, $^{115}$In, $^{69}$Ga and $^{71}$Ga. 

For a specific $I$, $\nu_{Q}$ and $\nu_{RF}$, we occupy the initial nuclear states according to the known $T$, and integrate the Schr\"{o}dinger equation numerically to determine $\left<I_{z}\right>$ after a single NMR pulse, converting $\left<I_{z}\right>$ to $\Delta_{\rm OHS}$ with the appropriate hyperfine coefficient. We find that the $\nu_{2}$-dependence is a strong function of both $\left< \nu_Q \right>$ and $p(\nu_{Q})$ \cite{suppl} and is therefore ideal to determine them. The $^{75}$As and $^{71}$Ga are well isolated as a function of $\nu_{2}$ and in both cases, $\left< \nu_Q \right>$ and $p(\nu_{Q})$ are readily determined by comparing the experimental results to the theory. The $^{69}$Ga $\nu_{2}$-dependence can be predicted from the $^{71}$Ga $\nu_{2}$-dependence simply by the known abundances and quadrupole moments \cite{suppl}. The remaining signal at intermediate $\nu_{2}$ arises mostly from $^{115}$In allowing us to determine the $^{115}$In quadrupole parameters. Fig.\ 4 shows that, first, we achieve an excellent  description of the experimental results; and second, the signals from the four isotopes $^{75}$As, $^{115}$In, $^{69}$Ga and $^{71}$Ga overlap little facilitating the determination of each quadrupole distribution.

We return to the sweep rate dependence. We calculate the $\alpha$-dependence, adding the results from each isotope with $x$, $T$, $\left< \nu_Q \right>$ as input parameters. ($B_{x}$ is adjusted within its error window to ensure that the plateau occurs at the correct $\alpha$.) The {\em same} set of parameters describes {\em both} the $\nu_{2}$- and $\alpha$-dependences. Fig.\ 3 shows the contribution from each isotope. $^{115}$In has the largest $h \nu_{\text{eff}}$ (on account of its large spin, $I=\frac{9}{2}$) and inversion at the first quantum transition is achieved first of all, closely followed by inversion at the first quantum transition for the $I=\frac{3}{2}$ nuclei. At the smallest $\alpha$, inversion at the second quantum transition is achieved for most of the In nuclei (and some of the $^{71}$Ga nuclei) but for most of the $^{75}$As and $^{69}$Ga nuclei, inversion at the first quantum transition is complete but inversion at the second quantum transition is not yet achieved. This explains the second change in gradient at the smallest $\alpha$ in the experiment. The combination of the $\nu_{2}$ and the $\alpha$-dependences allows in principle an initial nuclear spin temperature to be determined for each isotope. In practice, these temperatures are not significantly different to within the random error \cite{suppl} and we take a common temperature for simplicity.

The overall conclusion is that frequency-swept NMR enables the determination of all key parameters of the nuclear spins even at the single quantum dot level: the chemical composition, the effective temperatures and the quadrupole frequency distribution of each isotope.

As an outlook, we note that a sweep adiabatic for $|\Delta m=1|$ but sudden for $|\Delta m=2|$ can be used to produce highly non-thermal distributions of the spin states, boosting the NMR signal of the central transitions. Also, at an intermediate sweep rate, a superposition of the spin states is created with a chirped NMR pulse, and back-and-forth frequency sweeps result in quantum interferences, the St\"{u}ckelberg oscillations \cite{Stuckelberg1932, Yoakum1992, Oliver2005, Huang2011, Shevchenko2010}. This experiment represents the ideal springboard to explore quantum coherence in a complex nuclear spin ensemble using multiple chirped pulses.

\noindent{\bf Acknowledgements}\\
M.M., G.W., A.V.K., M.P. and R.J.W. acknowledge support from NCCR QSIT and EU ITN S$^{3}$NANO; M.P. and F.X. from the SNI; A.L., D.R. and A.D.W. from Mercur Pr-2013-0001 and BMBF-Q.com-H 16KIS0109. We thank Phani Peddibhotla and Christoph Kloeffel for technical assistance and Christian Degen, Patrick Maletinsky and Hugo Ribeiro for fruitful discussions.\\

\noindent{\bf Author contributions}\\
M.M. and G.W. carried out the experiments, the data analysis and the theoretical modeling. A.V.K. provided expertise in resonance fluorescence on single quantum dots; F.X. expertise in micro-wire design and sample processing. F.X. and M.P. provided electronics and software expertise for the NMR. A.L., D.R. and A.D.W. carried out the molecular beam epitaxy. M.M., G.W., M.P. and R.J.W. took the lead in writing the paper/supplementary information. R.J.W. conceived and managed the project.\\

\noindent{\bf Additional information}\\
Supplementary information is available in the online version of the paper. Reprints and permissions information is available online at www.nature.com/reprints. Correspondence and requests for materials should be addressed to M.M.\\

\noindent{\bf Competing financial interests}
The authors declare no competing financial interests.

\newpage

\begin{figure}[htb]
\includegraphics[width=\textwidth]{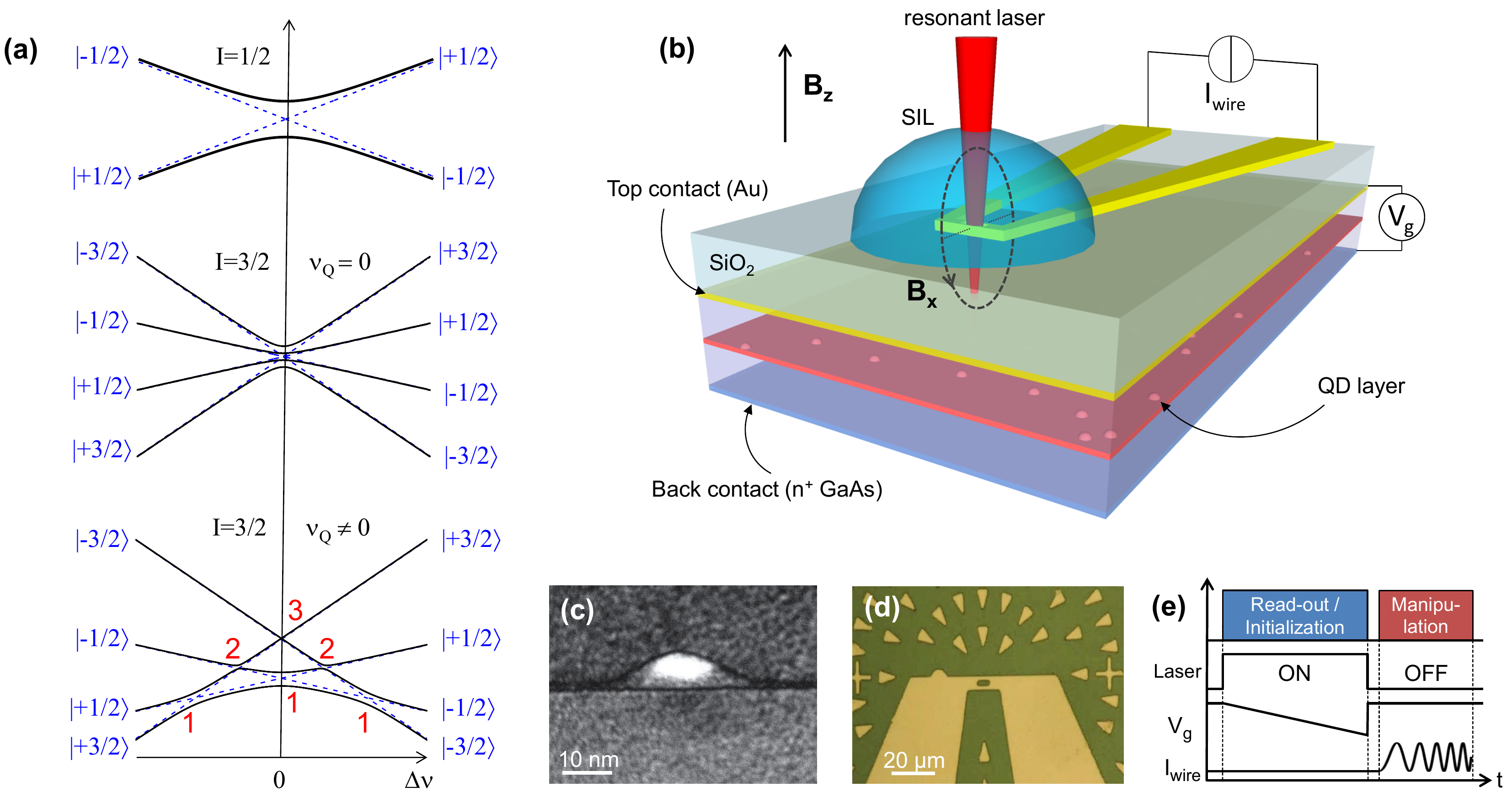}
\caption{\textbf{The experiment: concepts and design.} (a) Eigen-energies of the nuclear spin in a static magnetic field along $z$ and oscillating (radio-frequency, RF) magnetic field along $x$, in the rotating frame. In black are the eigen-energies versus RF detuning in three cases. In blue are the diabatic states. Top: $I=\frac{1}{2}$, a two-level system with avoided-crossing at $\Delta \nu=0$, $\ket{\uparrow} \equiv \ket{+\frac{1}{2}}\, , \,\ket{\downarrow} \equiv \ket{-\frac{1}{2}}$; middle: $I=\frac{3}{2}$ without quadrupole interaction; bottom: $I=\frac{3}{2}$ with quadrupole interaction ($ \nu_{Q} \gg \nu_{RF}$) showing a hierarchy of avoided crossings, the first, second and third quantum transitions ($|\Delta m=1|$, $|\Delta m=2|$ and $|\Delta m=3|$, respectively) \cite{suppl}. (b) Device for magnetic resonance experiments on the nuclear spins of a single self-assembled quantum dots. The quantum dots are embedded in a vertical tunnelling structure controlled by gate voltage $V_{g}$. A gold microwire is fabricated above the gate with a hole for optical access. Magnetic resonance is driven with an RF current passing through the microwire. A solid-immersion-lens enhances the collection efficiency of the resonance fluorescence. (c) Cross-section of a single InGaAs quantum dot (TEM image courtesy of Arne Ludwig and Jean-Michel Chauveau). (d) Top view of microwire. (e) Pulse sequence of NMR experiment. A resonance is established with a constant frequency laser. On ramping the gate voltage, the nuclear spins polarize in order to maintain the optical resonance: the Stark effect is compensated by the Overhauser shift. A RF pulse is then applied to manipulate the nuclear spin ensemble. The optical sequence is repeated to read-out the nuclear spin polarization, acting also as initialization for the next sequence.}
\label{Fig1}
\end{figure} 
\begin{figure}[htb]
\includegraphics[width=\textwidth]{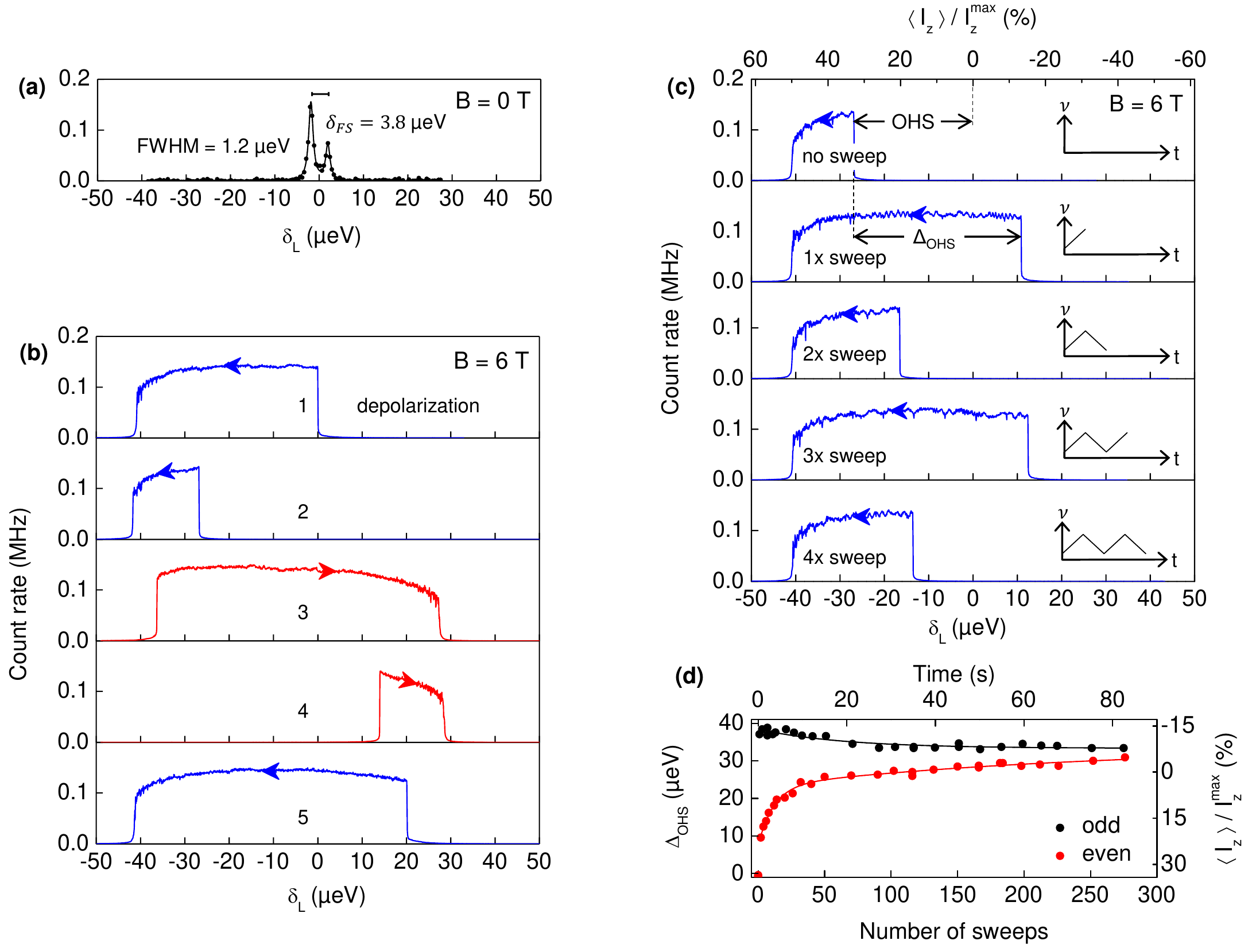}
\caption{\textbf{Adiabatic passage of the nuclear spin ensemble.} (a) Resonance fluorescence versus laser detuning on an empty single quantum dot (X$^{0}$ transition) at $B_{z}=0$\,T and $T=4$\,K. (b) Resonance fluorescence versus laser detuning at $B_{z}=6$\,T on the blue X$^{0}$ transition showing ``dragging". The plateau-like features signify nuclear spin polarization. A sequence of sweeps shows clear memory effects. The extent of the plateaux are reproducible to within 0.6 \textmu eV on repeating a specific cycle. In blue (red) the laser is tuned to more negative (positive) values. (c) A sequence of resonance fluorescence sweeps with $N$ chirped RF pulses ($\nu_1=32.5$\,MHz, $\nu_2=87.5$\,MHz, $\alpha=0.18$\,GHz/s) following nuclear spin polarization ($N=0,1,2,3,4$). $N=0$ reads initial $\left< I_{z} \right>$ \cite{suppl}; $N=1$ inverts $\left< I_{z} \right>$; $N=2$ restores $\left< I_{z} \right>$ to almost its $N=0$ value, etc. The Overhauser shift (OHS) and the change in Overhauser shift $\Delta_{\rm OHS}$  following a chirped pulse are labelled. (d) $\Delta_{\rm OHS}$ versus $N$ for large $N$. The decay at large $N$ arises mostly from relaxation processes during the sweep; the residual signal at large $N$ is presently not understood. Solid lines are guides for the eye.}
\label{Fig2}
\end{figure}
\begin{figure}[htb]
\includegraphics[width=0.8\textwidth]{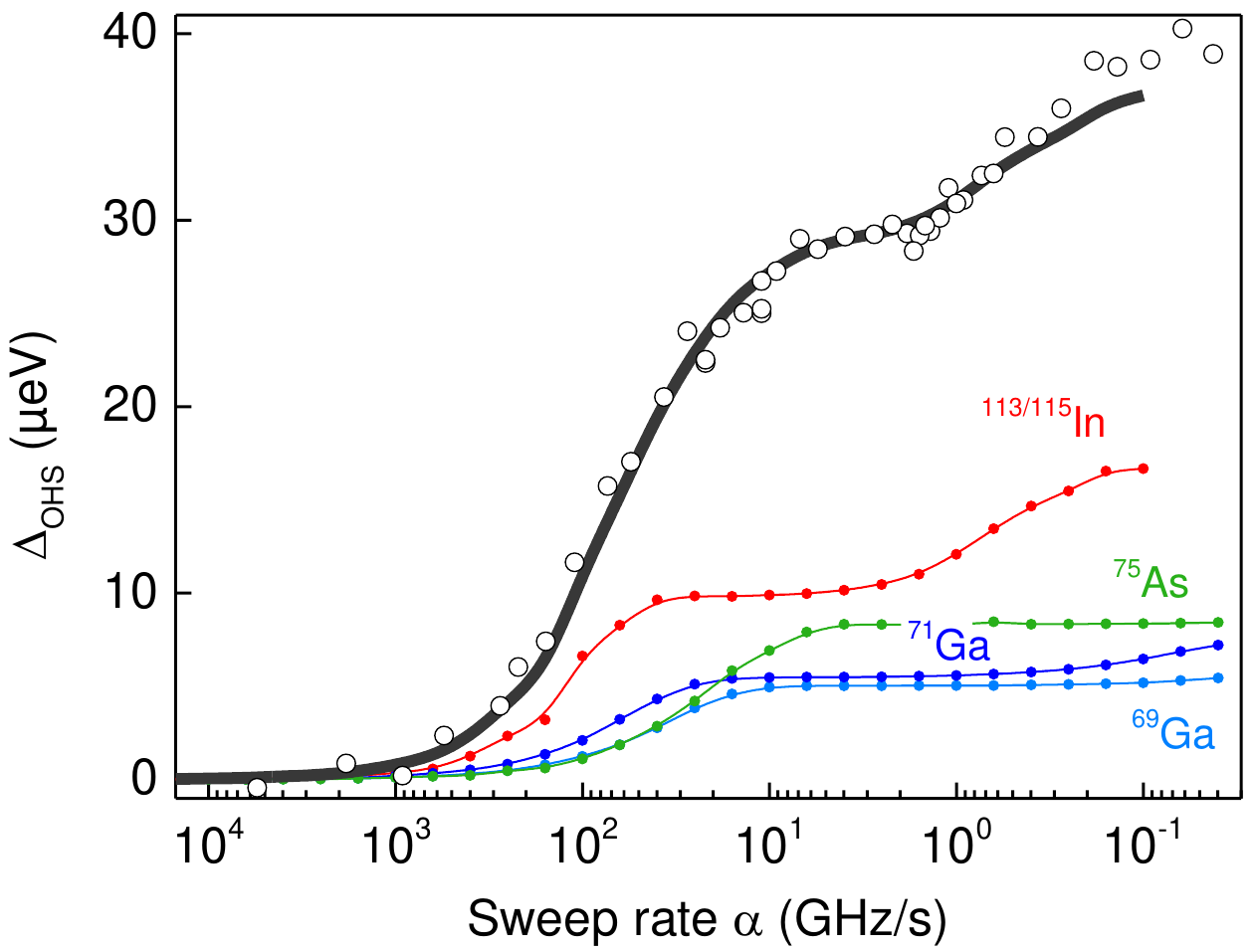}
\caption{\textbf{Nuclear spin inversion at the first quantum transition in chirped NMR.}  NMR signal, $\Delta_{\rm OHS}$, following a single chirped RF pulse with $\nu_{1} \rightarrow \nu_{2}$ ($\nu_{1}=32.5$ MHz, $\nu_{2}=87.5$ MHz) as a function of sweep rate $\alpha$: experimental data (open circles) along with theory (dark gray line). The theory uses $x=20.2$\%, $T=8.2$ mK, $B_{x}=3.8$ mT, $\left<\nu_{Q}[^{75}{\rm As}]\right>=3.0$ MHz, $\left<\nu_{Q}[^{115}{\rm In}]\right>=1.5$ MHz, $\left<\nu_{Q}[^{69}{\rm Ga}]\right>=3.1$ MHz, $\left<\nu_{Q}[^{71}{\rm Ga}]\right>=2.1$ MHz. The relative abundances are $^{75}$As (100\%), $^{113}$In (4.3\%), $^{115}$In (95.7\%); $^{69}$Ga (60.1\%) $^{71}$Ga (39.9\%). $\Delta_{\rm OHS}$ versus $\alpha$ is shown for the four isotopes separately (colour plots). The plateau arises because a range of $\alpha$ exists in which inversion at the first quantum transition is achieved for all isotopes yet inversion at the second quantum transition is achieved for none. At the smallest $\alpha$, inversion at the first and second quantum transitions is achieved for the majority of In nuclei but only inversion at the first quantum transition for the majority of $I=\frac{3}{2}$ nuclei.}
\label{Fig3}
\end{figure}
\begin{figure}[hbt]
\includegraphics[width=0.8\textwidth]{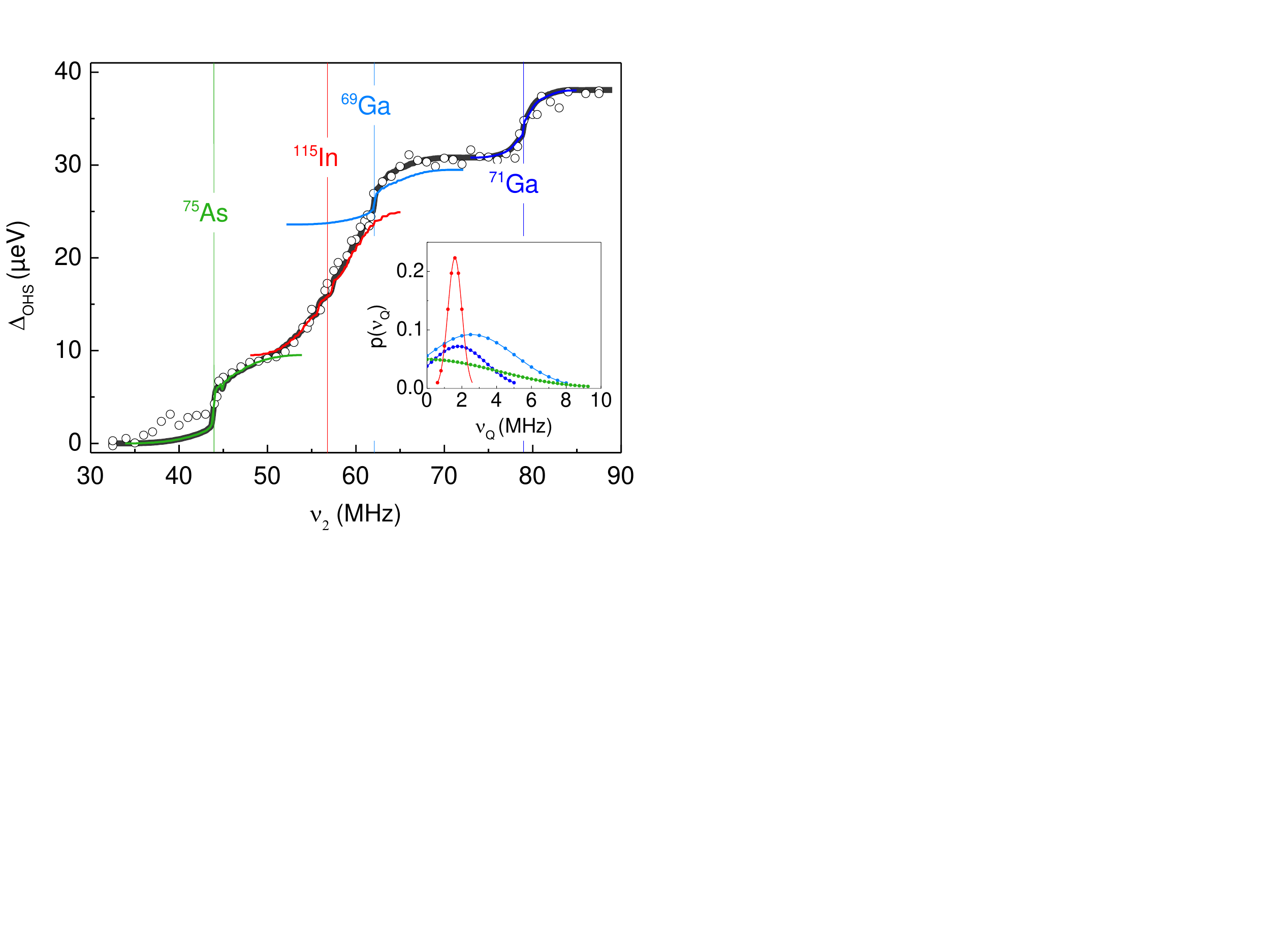}
\caption{\textbf{Isotope-sensitive NMR with chirped pulses.} NMR signal $\Delta_{\rm OHS}$ following a single chirped RF pulse, $\nu_{1} \rightarrow \nu_{2}$ as a function of $\nu_{2}$: experimental data (open circles) along with theory (dark gray line). The sweep rate $\alpha=0.09$ GHz/s and $\nu_{1}=32.5$ MHz. The vertical lines show the text-book NMR frequencies of the In ($I=\frac{9}{2}$), Ga ($I=\frac{3}{2}$) and As ($I=\frac{3}{2}$) isotopes: step-wise increases in signal occur each time $\nu_{2}$ crosses these particular frequencies. The theory uses $x=20.2$\%, $T=8.2$ mK and $B_{x}=3.8$ mT as in Fig.\ 3, along with Gaussian distributions for the quadrupole frequencies (inset).}
\label{Fig4}
\end{figure}


\begin{thebibliography}{10}
\expandafter\ifx\csname url\endcsname\relax
  \def\url#1{\texttt{#1}}\fi
\expandafter\ifx\csname urlprefix\endcsname\relax\def\urlprefix{URL }\fi
\providecommand{\bibinfo}[2]{#2}
\providecommand{\eprint}[2][]{\url{#2}}

\bibitem{Ribeiro2013}
\bibinfo{author}{Ribeiro, H.} \& \bibinfo{author}{Burkard, G.}
\newblock \bibinfo{title}{Nuclear spins keep coming back}.
\newblock \emph{\bibinfo{journal}{Nature Materials}}
  \textbf{\bibinfo{volume}{12}}, \bibinfo{pages}{469 -- 471}
  (\bibinfo{year}{2013}).

\bibitem{Warburton2013}
\bibinfo{author}{Warburton, R.~J.}
\newblock \bibinfo{title}{Single spins in self-assembled quantum dots}.
\newblock \emph{\bibinfo{journal}{Nature Materials}}
  \textbf{\bibinfo{volume}{12}}, \bibinfo{pages}{483 -- 493}
  (\bibinfo{year}{2013}).

\bibitem{Chekhovich2013}
\bibinfo{author}{Chekhovich, E.~A.} \emph{et~al.}
\newblock \bibinfo{title}{Nuclear spin effects in semiconductor quantum dots}.
\newblock \emph{\bibinfo{journal}{Nature Materials}}
  \textbf{\bibinfo{volume}{12}}, \bibinfo{pages}{494--504}
  (\bibinfo{year}{2013}).

\bibitem{Greilich2007}
\bibinfo{author}{Greilich, A.} \emph{et~al.}
\newblock \bibinfo{title}{Nuclei-induced frequency focusing of electron spin
  coherence}.
\newblock \emph{\bibinfo{journal}{Science}} \textbf{\bibinfo{volume}{317}},
  \bibinfo{pages}{1896--1899} (\bibinfo{year}{2007}).

\bibitem{Urbaszek2013}
\bibinfo{author}{Urbaszek, B.} \emph{et~al.}
\newblock \bibinfo{title}{Nuclear spin physics in quantum dots: An optical
  investigation}.
\newblock \emph{\bibinfo{journal}{Rev. Mod. Phys.}}
  \textbf{\bibinfo{volume}{85}}, \bibinfo{pages}{79--133}
  (\bibinfo{year}{2013}).

\bibitem{Chekhovich2012}
\bibinfo{author}{Chekhovich, E.} \emph{et~al.}
\newblock \bibinfo{title}{Structural analysis of strained quantum dots using
  nuclear magnetic resonance}.
\newblock \emph{\bibinfo{journal}{Nature Nanotechnology}}
  \textbf{\bibinfo{volume}{7}}, \bibinfo{pages}{646--650}
  (\bibinfo{year}{2012}).

\bibitem{Staudacher2013}
\bibinfo{author}{Staudacher, T.} \emph{et~al.}
\newblock \bibinfo{title}{Nuclear magnetic resonance spectroscopy on a
  (5-nanometer)3 sample volume}.
\newblock \emph{\bibinfo{journal}{Science}} \textbf{\bibinfo{volume}{339}},
  \bibinfo{pages}{561--563} (\bibinfo{year}{2013}).

\bibitem{Mamin2013}
\bibinfo{author}{Mamin, H.~J.} \emph{et~al.}
\newblock \bibinfo{title}{Nanoscale nuclear magnetic resonance with a
  nitrogen-vacancy spin sensor}.
\newblock \emph{\bibinfo{journal}{Science}} \textbf{\bibinfo{volume}{339}},
  \bibinfo{pages}{557--560} (\bibinfo{year}{2013}).

\bibitem{Maletinsky2009}
\bibinfo{author}{Maletinsky, P.}, \bibinfo{author}{Kroner, M.} \&
  \bibinfo{author}{Imamoglu, A.}
\newblock \bibinfo{title}{Breakdown of the nuclear-spin-temperature approach in
  quantum-dot demagnetization experiments}.
\newblock \emph{\bibinfo{journal}{Nature Physics}}
  \textbf{\bibinfo{volume}{5}}, \bibinfo{pages}{407--411}
  (\bibinfo{year}{2009}).

\bibitem{Latta2011b}
\bibinfo{author}{Latta, C.}, \bibinfo{author}{Srivastava, A.} \&
  \bibinfo{author}{Imamoglu, A.}
\newblock \bibinfo{title}{Hyperfine interaction-dominated dynamics of nuclear
  spins in self-assembled ingaas quantum dots}.
\newblock \emph{\bibinfo{journal}{Phys. Rev. Lett.}}
  \textbf{\bibinfo{volume}{107}}, \bibinfo{pages}{167401}
  (\bibinfo{year}{2011}).

\bibitem{suppl}
\bibinfo{note}{Supplementary Information}.

\bibitem{Gammon1997}
\bibinfo{author}{Gammon, D.} \emph{et~al.}
\newblock \bibinfo{title}{Nuclear spectroscopy in single quantum dots:
  Nanoscopic raman scattering and nuclear magnetic resonance}.
\newblock \emph{\bibinfo{journal}{Science}} \textbf{\bibinfo{volume}{277}},
  \bibinfo{pages}{85--88} (\bibinfo{year}{1997}).

\bibitem{Makhonin2011}
\bibinfo{author}{Makhonin, M.} \emph{et~al.}
\newblock \bibinfo{title}{Fast control of nuclear spin polarization in an
  optically pumped single quantum dot}.
\newblock \emph{\bibinfo{journal}{Nature Materials}}
  \textbf{\bibinfo{volume}{10}}, \bibinfo{pages}{844--848}
  (\bibinfo{year}{2011}).

\bibitem{Flisinski2010}
\bibinfo{author}{Flisinski, K.} \emph{et~al.}
\newblock \bibinfo{title}{Optically detected magnetic resonance at the
  quadrupole-split nuclear states in (in,ga)as/gaas quantum dots}.
\newblock \emph{\bibinfo{journal}{Phys. Rev. B}} \textbf{\bibinfo{volume}{82}},
  \bibinfo{pages}{081308} (\bibinfo{year}{2010}).

\bibitem{Cherbunin2011}
\bibinfo{author}{Cherbunin, R.~V.} \emph{et~al.}
\newblock \bibinfo{title}{Resonant nuclear spin pumping in (in,ga)as quantum
  dots}.
\newblock \emph{\bibinfo{journal}{Phys. Rev. B}} \textbf{\bibinfo{volume}{84}},
  \bibinfo{pages}{041304} (\bibinfo{year}{2011}).

\bibitem{Bulutay2012}
\bibinfo{author}{Bulutay, C.}
\newblock \bibinfo{title}{Quadrupolar spectra of nuclear spins in strained
  ingaas quantum dots}.
\newblock \emph{\bibinfo{journal}{Phys. Rev. B}} \textbf{\bibinfo{volume}{85}},
  \bibinfo{pages}{115313} (\bibinfo{year}{2012}).

\bibitem{Peddibhotla2013}
\bibinfo{author}{Peddibhotla, P.} \emph{et~al.}
\newblock \bibinfo{title}{Harnessing nuclear spin polarization fluctuations in
  a semiconductor nanowire}.
\newblock \emph{\bibinfo{journal}{Nature Physics}}
  \textbf{\bibinfo{volume}{9}}, \bibinfo{pages}{631} (\bibinfo{year}{2013}).

\bibitem{Shevchenko2010}
\bibinfo{author}{Shevchenko, S.}, \bibinfo{author}{Ashhab, S.} \&
  \bibinfo{author}{Nori, F.}
\newblock \bibinfo{title}{Landau-{Z}ener-{S}t{\"u}ckelberg interferometry}.
\newblock \emph{\bibinfo{journal}{Physics Reports}}
  \textbf{\bibinfo{volume}{492}}, \bibinfo{pages}{1--30}
  (\bibinfo{year}{2010}).

\bibitem{Poggio2007}
\bibinfo{author}{Poggio, M.}, \bibinfo{author}{Degen, C.~L.},
  \bibinfo{author}{Rettner, C.}, \bibinfo{author}{Mamin, H.} \&
  \bibinfo{author}{Rugar, D.}
\newblock \bibinfo{title}{Nuclear magnetic resonance force microscopy with a
  microwire rf source}.
\newblock \emph{\bibinfo{journal}{Applied Physics Letters}}
  \textbf{\bibinfo{volume}{90}}, \bibinfo{pages}{263111}
  (\bibinfo{year}{2007}).

\bibitem{Kuhlmann2013a}
\bibinfo{author}{Kuhlmann, A.~V.} \emph{et~al.}
\newblock \bibinfo{title}{Charge noise and spin noise in a semiconductor
  quantum device}.
\newblock \emph{\bibinfo{journal}{Nature Physics}}
  \textbf{\bibinfo{volume}{9}}, \bibinfo{pages}{570--575}
  (\bibinfo{year}{2013}).

\bibitem{Kuhlmann2013b}
\bibinfo{author}{Kuhlmann, A.~V.} \emph{et~al.}
\newblock \bibinfo{title}{A dark-field microscope for background-free detection
  of resonance fluorescence from single semiconductor quantum dots operating in
  a set-and-forget mode}.
\newblock \emph{\bibinfo{journal}{Review of Scientific Instruments}}
  \textbf{\bibinfo{volume}{84}}, \bibinfo{pages}{073905}
  (\bibinfo{year}{2013}).

\bibitem{Kuhlmann2013c}
\bibinfo{author}{Kuhlmann, A.~V.} \emph{et~al.}
\newblock \bibinfo{title}{Linewidth of single photons from a single quantum
  dot: key role of nuclear spins}.
\newblock \emph{\bibinfo{journal}{arXiv:1307.7109}}  (\bibinfo{year}{2013}).

\bibitem{Latta2009}
\bibinfo{author}{Latta, C.} \emph{et~al.}
\newblock \bibinfo{title}{Confluence of resonant laser excitation and
  bidirectional quantum-dot nuclear-spin polarization}.
\newblock \emph{\bibinfo{journal}{Nature Physics}}
  \textbf{\bibinfo{volume}{5}}, \bibinfo{pages}{758--763}
  (\bibinfo{year}{2009}).

\bibitem{Hogele2012}
\bibinfo{author}{H\"ogele, A.} \emph{et~al.}
\newblock \bibinfo{title}{Dynamic nuclear spin polarization in the resonant
  laser excitation of an ingaas quantum dot}.
\newblock \emph{\bibinfo{journal}{Phys. Rev. Lett.}}
  \textbf{\bibinfo{volume}{108}}, \bibinfo{pages}{197403}
  (\bibinfo{year}{2012}).

\bibitem{Vega1978}
\bibinfo{author}{Vega, S.}
\newblock \bibinfo{title}{Fictitious spin 1/2 operator formalism for multiple
  quantum {NMR}}.
\newblock \emph{\bibinfo{journal}{The Journal of Chemical Physics}}
  \textbf{\bibinfo{volume}{68}}, \bibinfo{pages}{5518--5527}
  (\bibinfo{year}{1978}).

\bibitem{Haase1994}
\bibinfo{author}{Haase, J.}, \bibinfo{author}{Conradi, M.},
  \bibinfo{author}{Grey, C.} \& \bibinfo{author}{Vega, A.}
\newblock \bibinfo{title}{Population transfers for {NMR} of quadrupolar spins
  in solids}.
\newblock \emph{\bibinfo{journal}{Journal of Magnetic Resonance, Series A}}
  \textbf{\bibinfo{volume}{109}}, \bibinfo{pages}{90--97}
  (\bibinfo{year}{1994}).

\bibitem{Veenendaal1998}
\bibinfo{author}{van Veenendaal, E.}, \bibinfo{author}{Meier, B.~H.} \&
  \bibinfo{author}{Kentgens, A. P.~M.}
\newblock \bibinfo{title}{Frequency stepped adiabatic passage excitation of
  half-integer quadrupolar spin systems}.
\newblock \emph{\bibinfo{journal}{Molecular Physics}}
  \textbf{\bibinfo{volume}{93}}, \bibinfo{pages}{195--213}
  (\bibinfo{year}{1998}).

\bibitem{Stuckelberg1932}
\bibinfo{author}{St\"{u}ckelberg, E.}
\newblock \bibinfo{title}{Theorie der unelastischen {S}t\"{o}ssen zwischen
  {A}tomen}.
\newblock \emph{\bibinfo{journal}{Helv. Phys. Acta.}}
  \textbf{\bibinfo{volume}{5}}, \bibinfo{pages}{369} (\bibinfo{year}{1932}).

\bibitem{Yoakum1992}
\bibinfo{author}{Yoakum, S.}, \bibinfo{author}{Sirko, L.} \&
  \bibinfo{author}{Koch, P.~M.}
\newblock \bibinfo{title}{Stueckelberg oscillations in the multiphoton
  excitation of helium rydberg atoms: Observation with a pulse of coherent
  field and suppression by additive noise}.
\newblock \emph{\bibinfo{journal}{Phys. Rev. Lett.}}
  \textbf{\bibinfo{volume}{69}}, \bibinfo{pages}{1919--1922}
  (\bibinfo{year}{1992}).

\bibitem{Oliver2005}
\bibinfo{author}{Oliver, W.~D.} \emph{et~al.}
\newblock \bibinfo{title}{Mach-{Z}ehnder interferometry in a strongly driven
  superconducting qubit}.
\newblock \emph{\bibinfo{journal}{Science}} \textbf{\bibinfo{volume}{310}},
  \bibinfo{pages}{1653--1657} (\bibinfo{year}{2005}).

\bibitem{Huang2011}
\bibinfo{author}{Huang, P.} \emph{et~al.}
\newblock \bibinfo{title}{Landau-{Z}ener-{S}t\"uckelberg interferometry of a
  single electronic spin in a noisy environment}.
\newblock \emph{\bibinfo{journal}{Phys. Rev. X}} \textbf{\bibinfo{volume}{1}},
  \bibinfo{pages}{011003} (\bibinfo{year}{2011}).

\end{thebibliography}
\end{document}